\newcommand{\be}{\begin{equation}}
\newcommand{\ee}{\end{equation}}
\newcommand{\bea}{\begin{eqnarray}}
\newcommand{\eea}{\end{eqnarray}}
\newcommand{\nn}{\nonumber}
\newcommand{\Appendix}[1]%
    {\renewcommand{\thesection}{Appendix~\Alph{section}:}%
         \section{#1}}%
\catcode`@=11
\long\def\@makecaption#1#2{
   \vskip 10pt
   \setbox\@tempboxa\hbox{{\small\bf #1.} \ {\small #2}}
   \ifdim \wd\@tempboxa >\hsize       % IF longer than one line:
   {\small\bf #1.} \ {\small #2}\par  % THEN set as ordinary paragraph.
   \else                              %   ELSE  center.
        \hbox to\hsize{\hfil\box\@tempboxa\hfil}
   \fi}
\catcode`@=12

\catcode`@=11
\def\secteqno{\@addtoreset{equation}{section}%
\def\theequation{\thesection.\arabic{equation}}}
\def\endsecteqno{\def\theequation{\@ifundefined{chapter}%
{\arabic{equation}}{\thechapter.\arabic{equation}}}}
\newcounter{subequation}
\def\thesubequation{\alph{subequation}}
\def\sneqnarray{\stepcounter{equation}\let\@currentlabel=\theequation
\setcounter{subequation}{1}
\def\@eqnnum{{\rm (\theequation\thesubequation)}}
\global\@eqcnt\z@\tabskip\@centering\let\\=\@eqncr\let\@@eqncr=\@@sneqncr
$$\halign to \displaywidth\bgroup\@eqnsel\hskip\@centering
 $\displaystyle\tabskip\z@{##}$&\global\@eqcnt\@ne
 \hskip 2\arraycolsep \hfil${##}$\hfil
 &\global\@eqcnt\tw@ \hskip 2\arraycolsep
$\displaystyle\tabskip\z@{##}$\hfil
tabskip\@centering&\llap{##}\tabskip\z@\cr}
\def\endsneqnarray{\@@sneqncr\egroup $$\global\@ignoretrue}
\def\@@sneqncr{\let\@tempa\relax
   \ifcase\@eqcnt \def\@tempa{& & &}\or \def\@tempa{& &}
   \else \def\@tempa{&}\fi
     \@tempa \if@eqnsw\@eqnnum\stepcounter{subequation}\fi
     \global\@eqnswtrue\global\@eqcnt\z@\cr}
\def\nobiblabels{\def\@lbibitem[##1]##2{\@bibitem{##2}}}
\catcode`@=12

\def\beq{\begin{equation}}
\def\eeq{\end{equation}}
\def\bea{\begin{eqnarray}}
\def\eea{\end{eqnarray}}
\def\nn{\nonumber}

\def\lQ{\Lambda_{\rm QCD}}

\def\bnabla{{\bm \nabla}}

\documentclass[aps,10pt,prd,showpacs,amsmath,amssymb,preprintnumbers,superscriptaddress,nofootinbib,showkeys]{revtex4-2}

\usepackage[german,english]{babel}
\usepackage{hyperref}
\usepackage{amssymb}
\usepackage{amsmath}
\usepackage{bm}% bold math   
\usepackage{braket}
\usepackage{slashed}
\usepackage{multirow}
\usepackage{epsfig}
\usepackage{epstopdf}
\usepackage{bbm}
\usepackage{bbold}
\usepackage[mathlines]{lineno}% Enable numbering of text and display math
\usepackage{soul}%remove eventually

\begin{document}

\title{Nonrelativistic effective field theory for heavy exotic hadrons}

\author{Joan Soto}
\email{joan.soto@ub.edu}
\affiliation{Departament de F\'\i sica Qu\`antica i Astrof\'isica and Institut de Ci\`encies del Cosmos, Universitat de Barcelona, Mart\'\i$\,$ i Franqu\`es 1, 08028 Barcelona, Catalonia, Spain}

\author{Jaume Tarr\'us Castell\`a}
\email{jtarrus@ifae.es}
\affiliation{Grup de F\'\i sica Te\`orica, Dept. F\'\i sica and IFAE-BIST, Universitat Aut\`onoma de Barcelona,\\ 
E-08193 Bellaterra (Barcelona), Catalonia, Spain}

\date{\today}

\begin{abstract}
We propose an effective field theory to describe hadrons with two heavy quarks without any assumption on the typical distance between the heavy quarks with respect to the typical hadronic scale. The construction is based on nonrelativistic QCD and inspired in the strong coupling regime of potential nonrelativistic QCD. We construct the effective theory at leading and next-to-leading order in the inverse heavy quark mass expansion for arbitrary quantum numbers of the light degrees of freedom. Hence our results hold for hybrids, tetraquarks, double heavy baryons and pentaquarks, for which we also present the corresponding operators at nonrelativistic level. At leading order, the effective theory enjoys heavy quark spin symmetry and corresponds to the Born-Oppenheimer approximation. At next-to-leading order, spin and velocity-dependent terms arise, which produce splittings in the heavy quark spin symmetry multiplets. A concrete application to double heavy baryons is presented in an accompanying paper.
\end{abstract}

\maketitle

\section{Introduction} 

Exotic hadrons, which may be defined as hadrons that are neither mesons (quark-antiquark states) nor baryons (three quark states), were already foreseen in the early days of QCD~\cite{Jaffe:1975fd}. Whereas mesons and baryons are well-defined objects in the nonrelativistic quark model, they are not so in the context of QCD. This is because the light quark masses are much smaller than the typical hadronic scale $\lQ$, and hence light quark-antiquark pairs can be easily created both in mesons and baryons turning them into tetraquark or pentaquark states, or even into states with a higher number of quarks and antiquarks. The situation changes dramatically in the case of hadrons containing heavy quarks, such as the charm and bottom. Since their masses are much larger than $\lQ$, the creation of heavy quark-antiquark pairs in hadrons is highly suppressed. Therefore, the number of heavy quarks in a heavy hadron is closely related to its mass, and the quantum numbers such as isospin and baryon number indicate the light quark content. 

For the last two decades numerous experimental collaborations have discovered and measured the properties of many unexpected hadrons, which have been generically named as $XYZ$. The first of these states discovered was the $X(3872)$~\cite{Choi:2003ue}, which turns out to be extremely close to the $D^0$-$D^{0\ast}$ meson threshold, and hence a molecular interpretation is tempting. Later on, another charmoniumlike state was found, the $Y(4260)$~\cite{Aubert:2005rm}, whose weak production in $e^+e^-$ annihilation signals it as an exotic. In the following years charged states in the charmonium and bottomonium spectrum where discovered~\cite{Choi:2007wga,Belle:2011aa,Aaij:2014jqa} which can be unambiguously identified as tetraquarks. More recently, at the LHCb experiment, two isospin $1/2$ baryons where found with masses close to the charmonium states~\cite{Aaij:2015tga} which have been interpreted as a pentaquark formed by a charm-anticharm pair and three light quarks. More data from experiments is expected in the near future due to the ongoing experimental efforts at facilities such as BESIII, Belle II, LHC and in the future at FAIR, which stresses the need for a comprehensive theoretical framework based on QCD to describe these states. 

Heavy quarkonium, namely heavy quark-antiquark systems, have also been studied since the early days of QCD, and in fact they played an important role in the consolidation of this theory~\cite{Appelquist:1974zd,DeRujula:1974rkb}. Since the heavy quark masses $m_Q$ are much larger than $\lQ$, it was soon realized that heavy quarks move slowly and hence they could be described by standard nonrelativistic quantum mechanics once the interaction potential was obtained from QCD, and this could be done in terms of the expectation value of the Wilson loop~\cite{Wilson:1974sk}. Corrections to the leading-order mass-independent potential were also considered and expressed again in terms of expectation values of operator insertions in the Wilson loop~\cite{Eichten:1980mw,Gromes:1984ma,Barchielli:1986zs,Barchielli:1988zp}. All these developments were later on recast in the framework of nonrelativistic effective field theories~\cite{Caswell:1985ui,Pineda:1997bj} (see Ref.~\cite{Brambilla:2004jw} for a review), which allowed one to incorporate the so-called hard corrections~\cite{Chen:1994dg,Pineda:1998kj}, and provided a complete result for the potential at order $1/m_Q^2$~\cite{Brambilla:2000gk,Pineda:2000sz}. The different terms in the potential have been evaluated in lattice QCD~\cite{Bali:1997am,Koma:2006si,Koma:2006fw}. 

It is the aim of this paper to put forward a general effective field theory (EFT) framework, analogous to the one described above for heavy quarkonium, for any heavy hadron, containing a heavy quark-antiquark or two heavy quarks, and a gluon and light quark state with arbitrary quantum numbers. We shall refer to the gluon and light quarks collectively as light degrees of freedom (LDF). We will call heavy exotic hadron to any such state with LDF quantum numbers different from $0^{++}$, this case corresponding to heavy quarkonium. Note that, for convenience, we include double heavy baryons in this definition of heavy exotic hadron\footnote{Note also that we exclude $0^{++}$ excitations of the LDF, which indeed exist in the spectrum \cite{Juge:1999ie}. In that case, the formulas for the $1/m_Q$ corrections are analogous to the ones for heavy quarkonium case \cite{Brambilla:2000gk,Pineda:2000sz}. In particular, no spin and velocity dependent corrections appear at order $1/m_Q$. This is also true in the more general case of $0^\pm$ LDF.}. Heavy exotic hadrons are then composed of two distinct components: the heavy quarks and the LDF. The former form a nonrelativistic bound state with an interaction potential depending on the LDF state. Such bound states are characterized by three well-separated scales: the heavy quark mass, $m_Q$, the relative momentum $m_Q v$, with $v\ll 1$ the relative velocity, and the binding energy $m_Q v^2$. The LDF states are characterized by the typical hadronic scale $\lQ$. The EFT that we present in this paper is built from QCD using two energy gaps between these characteristic scales. The first one is the aforementioned fact that $m_Q\gg\lQ$, which is implemented in an EFT framework in nonrelativistic QCD (NRQCD)~\cite{Caswell:1985ui,Bodwin:1994jh}. The second one is $\lQ\gg m_Q v^2$, which is nothing else than the observation that one can perform an adiabatic expansion between the dynamics of the heavy degrees of freedom, the heavy quarks, and the LDF, the gluons and light quarks. Furthermore, in the short-distance regime, $m_Q v\sim 1/r\gg\lQ$, the relative momentum scale can be integrated out perturbatively. This is the so-called weak-coupling regime of potential NRQCD (pNRQCD)~\cite{Pineda:1997bj,Brambilla:1999xf} developed originally for quark-antiquark systems to study standard quarkonium and later on adapted to quark-quark systems to study double heavy baryons~\cite{Brambilla:2005yk,Mehen:2019cxn}.

The leading order in the adiabatic expansion is the celebrated Born-Oppenheimer approximation, which has been applied to heavy hybrids for quite some time in Refs.~\cite{Griffiths:1983ah,Juge:1999ie,Guo:2008yz,Braaten:2014qka,Capitani:2018rox,Mueller:2019mkh} and put into a nonrelativistic EFT framework in Refs.~\cite{Berwein:2015vca,Brambilla:2017uyf,Oncala:2017hop}. The Born-Oppenheimer approximation has also been used for heavy tetraquarks in Ref.~\cite{Braaten:2014qka,Bicudo:2015kna,Bicudo:2016ooe,Bicudo:2017szl,Bicudo:2019ymo}. The EFT for heavy hybrids has been extended beyond leading order to include spin-dependent operators up to $1/m_Q$~\cite{Soto:2017one} and up to $1/m^2_Q$~\cite{Brambilla:2018pyn,Brambilla:2019jfi}. In Refs.~\cite{Berwein:2015vca,Brambilla:2018pyn,Brambilla:2019jfi} the dependence of the potentials in the interquark distance, $r$, was obtained in the short-distance regime, that is, assuming that the typical size of the system $r$ is much smaller than the typical hadronic size $1/\lQ$. In the case of the static potential, which can be obtained from the lattice QCD calculations, the results from Ref.~\cite{Berwein:2015vca} show that the short-distance regime is a reasonable approximation only for the lowest lying state in the charmonium sector and for a few low-lying states in the bottomonium one. This can be understood from a general standpoint from the fact that the hybrid static potential has a classical minimum at $r\sim 1/\lQ$ as was pointed out in Refs.~\cite{Oncala:2017hop,Soto:2017one}. It is not likely that the weak-coupling assumption is fulfilled in most cases for exotic hadrons for the actual values of the charm and bottom masses.

We shall then avoid any assumption on the relative size of $r$ and $1/\lQ$ here, and proceed in an analogous way to the strong coupling regime of pNRQCD~\cite{Brambilla:2000gk,Pineda:2000sz}, as it was advocated in Ref.~\cite{Brambilla:2008zz} and has already been initiated in the particular case of heavy hybrids~\cite{Oncala:2017hop,Soto:2017one}. The leading-order Lagrangian for the EFT for heavy exotic hadrons consists of wave function fields interacting with a mass-independent and heavy quark spin-independent potential, and coincides with the Born-Oppenheimer approximation. We consider here the complete $1/m_Q$ corrections containing the heavy quark spin or orbital angular momentum for an wide class of heavy exotic hadrons, including hybrids, tetraquarks and pentaquarks. We provide formulas for the potentials associated with the different operators in terms of expectation values of operator insertions in the Wilson loop, which are suitable to be calculated in lattice QCD. 

The paper is organized as follows. In Sec.~\ref{EFT}, we describe the EFT and write down the most general leading-order and next-to-leading order ($1/m_Q$) Lagrangian for arbitrary total angular momentum of LDF. In Sec.~\ref{match}, we describe the matching to NRQCD and provide formulas for the different terms in the potential that can be evaluated in lattice QCD. In Sec.~\ref{light}, we list the sets of LDF operators that are suitable to create hybrids, tetraquarks and pentaquarks. We close with Sec.~\ref{concl} devoted to discussion and conclusions. A concrete application of the general results in this paper to double heavy baryons is presented in an accompanying paper~\cite{Soto:2020pfa}. Some technical details are relegated to the Appendices~\ref{wlsimp} and \ref{tracecomp}.

\section{EFT for heavy exotic hadrons}
\label{EFT}

The EFT describing heavy exotic hadrons can be generically written as
\begin{align}
{\cal L}=&\sum_{\kappa^p}\Psi^{\dagger}_{\kappa^p}\left[i\partial_t-h_{\kappa^p}\right]\Psi_{\kappa^p}\,,\label{boeft}
\end{align}
where the sum over $\kappa^p$ refers to the sum over the LDF states with spin $\kappa$ and parity $p$ corresponding to the spectrum of static energies relevant for a given particular case. The LDF may have additional quantum numbers such as charge conjugation, isospin, strangeness or baryon number that we will not write explicitly. The $\Psi$ fields are understood as depending on $t,\,\bm{r},\,\bm{R}$, where $\bm{r}=\bm{x}_1-\bm{x}_2$ and $\bm{R}=(\bm{x}_1+\bm{x}_2)/2$ are the relative and center-of-mass coordinates of a heavy quark pair. The $\Psi$ fields live both in the LDF and heavy quark spin spaces. The field $\Psi^{\alpha}_{\kappa^p}$ corresponds to a spin $\kappa$ LDF state and has $\alpha=-\kappa,..,0,...\kappa$ components. In the Lagrangian in Eq.~\eqref{boeft} we have chosen to leave the spin indices implicit.

The Hamiltonian densities $h_{\kappa^p}$ have the following expansion up to $1/m_Q$
\begin{align}
h_{\kappa^p}=\frac{\bm{p}^2}{m_Q}+\frac{\bm{P}^2}{4m_Q}+V_{\kappa^p}^{(0)}(\bm{r})+\frac{1}{m_Q}V_{\kappa^p}^{(1)}(\bm{r},\,\bm{p})\,,\label{hamden}
\end{align}
with $\bm{p}=-i\bnabla_r$ and $\bm{P}=-i\bnabla_R$. The kinetic terms in Eq.~\eqref{hamden} are diagonal in spin space while the potentials are not. The static potentials, $V^{(0)}$, are diagonal in the heavy quark spin space, due to heavy quark spin symmetry, while the LDF spin structure is determined by the representations of $D_{\infty h}$ that the $\kappa^p$ quantum numbers can be associated with. $D_{\infty h}$ is a cylindrical symmetry group that characterizes any hadron composed of two heavy quarks. Its representations are labeled by $\Lambda$, the absolute value of the projection of the LDF state angular momentum on the axis joining the two heavy quarks\footnote{\label{cp} Additionally the representations are characterized by: $\eta=\pm 1$, the P or CP eigenvalue for heavy quark quark and heavy quark-antiquark systems, respectively, denoted by $g = + 1$ and $u = - 1$; and for $\Lambda=0$ there is a symmetry under reflection in any plane passing through the axis $\hat{\bm{r}}$, the eigenvalues of the corresponding symmetry operator being $\sigma=\pm 1$.}, $\hat{\bm{r}}$. Therefore $0\leq \Lambda \leq |\kappa|$. The static potential is expanded into $D_{\infty h}$ representations.
\begin{align}
V_{\kappa^p}^{(0)}(\bm{r})=&\,\sum_{\Lambda}V_{\kappa^p\Lambda}^{(0)}(r){\cal P}_{\kappa\Lambda}\,,\label{edihsp}
\end{align}
with ${\cal P}_{\kappa\Lambda}$ the projectors into representations of $D_{\infty h}$ in the spin-$\kappa$ space. These are $(2\kappa+1)\times (2\kappa+1)$ matrices in the light quark spin space and fulfill the usual projector properties: they are idempotent ${\cal P}^2_{\kappa\Lambda}={\cal P}_{\kappa\Lambda}$, orthogonal to each other ${\cal P}_{\kappa\Lambda}{\cal P}_{\kappa\Lambda'}=\delta_{\Lambda\Lambda'}$, and add up to the identity in the spin-$\kappa$ space $\sum_{\Lambda}{\cal P}_{\kappa\Lambda}=\mathbb{1}_{2\kappa+1}$.

The subleading potentials $V^{(1)}$ can be split into terms that depend on the total heavy quark spin, $\bm{S}_{QQ}$, or angular momentum, $\bm{L}_{QQ}$, and the terms independent of these two operators. The latter have the same structure as (\ref{edihsp}), whereas the former take the form
\begin{align}
V_{\kappa^p{\rm SD}}^{(1)}(\bm{r})=\sum_{\Lambda\Lambda'}{\cal P}_{\kappa\Lambda}\left[V^{sa}_{\kappa^p\Lambda\Lambda'}(r)\bm{S}_{QQ}\cdot\left({\cal P}^{\rm c.r.}_{10}\cdot\bm{S}_{\kappa}\right)+V^{sb}_{\kappa^p\Lambda\Lambda'}(r)\bm{S}_{QQ}\cdot\left({\cal P}^{\rm c.r.}_{11}\cdot\bm{S}_{\kappa}\right)+V^{l}_{\kappa^p\Lambda\Lambda'}(r)\left(\bm{L}_{QQ}\cdot\bm{S}_{\kappa}\right)\right]{\cal P}_{\kappa\Lambda'}\,.\label{sdp32}
\end{align}
The total heavy quark spin is defined as $2\bm{S}_{QQ}=\bm{\sigma}_{QQ}=\bm{\sigma}_{Q_1}\mathbb{1}_{2\,Q_2}+\mathbb{1}_{2\,Q_1}\bm{\sigma}_{Q_2}$ where the $\mathbb{1}_2$ are identity matrices in the heavy quark spin space for the heavy quark labeled in the subindex. The heavy quark angular momentum is defined as $\bm{L}_{QQ}=\bm{r}\times \bm{p}$. The superscript c.r. in the projectors indicates the use of the Cartesian basis representation of the spin-1 matrices, i.e. $\left(\bm{S}_1^i\right)^{jk}=-i\epsilon_{ijk}$. The heavy quark spin component of the $\Psi$ fields is given by $\chi_{s}^{Q_1}\chi_{r}^{Q_2}$ with $\chi_{s}$ the usual spin-$1/2$ two-component spinors. The LDF spin component, $\chi^\kappa_{\alpha}$, is a $(2\kappa+1)$-component spinor.

Let us summarize how to obtain the projectors ${\cal P}_{\kappa\Lambda}$ in any given spin-$\kappa$ space. First let us introduce the projection vectors $P_{\kappa\lambda}$ defined as the eigenvectors given by
\begin{align}
\left(\hat{\bm{r}}\cdot\bm{S}_{\kappa}\right)P_{\kappa\lambda}=\lambda\, P_{\kappa\lambda}\,.\label{eigeq}
\end{align}
These projection vectors for $\kappa=1$ were used in Refs.~\cite{Berwein:2015vca,Brambilla:2017uyf,Brambilla:2018pyn,Brambilla:2019jfi,Pineda:2019mhw} in the construction of EFT for hybrid quarkonium. The spin operator projected into the heavy quark pair axis can be expanded in these projector vectors
\begin{align}
\left(\hat{\bm{r}}\cdot\bm{S}_{\kappa}\right)=\sum_{\lambda}\lambda\, P_{\kappa\lambda}P^{\dagger}_{\kappa\lambda}\,.\label{pvex}
\end{align}
The projector into $D_{\infty h}$ representations in the spin-$\kappa$ space can then be defined as
\begin{align}
{\cal P}_{\kappa\Lambda}=\sum_{\lambda=\pm\Lambda}P_{\kappa\lambda}P^{\dagger}_{\kappa\lambda}\,.\label{defpro}
\end{align}
If we combine Eqs.~\eqref{pvex} and \eqref{defpro} we can arrive at
\begin{align}
\left(\hat{\bm{r}}\cdot\bm{S}_{\kappa}\right)^{2n}&=\sum_{\Lambda}\Lambda^{2n}{\cal P}_{\kappa\Lambda}\,.\label{niceeq}
\end{align}

Some interesting consequences follow from Eq.~\eqref{niceeq}. First, we can see that the Lagrangian in Eq.~\eqref{sdp32} contains all possible spin-dependent operators at order $1/m_Q$. The only way to construct new operators is to multiply the existing ones with scalar factors made out of $\hat{\bm{r}}$ and $\bm{S}_\kappa$. The only such scalar factors are precisely $\left(\hat{\bm{r}}\cdot\bm{S}_{\kappa}\right)^{2n}$ which due to Eq.~\eqref{niceeq} can be expanded in a linear combination of projectors. Finally, using the properties of the projectors the new operators can be reduced to the existing ones. The second interesting use of Eq.~\eqref{niceeq} is that it can be used to create a system of equations that can be inverted in order to obtain expressions for the projectors in terms of $\left(\hat{\bm{r}}\cdot\bm{S}_{\kappa}\right)^{2n}$ for $n=0,1,\dots$ up to the number of possible values of $0\leq\Lambda\leq|\kappa|$. In this way we obtain the projectors ${\cal P}$ up to spin-$2$
\begin{align}
{\cal P}_{\frac{1}{2}\frac{1}{2}}&=\mathbb{1}_{2}\,,\\
{\cal P}_{\frac{3}{2}\frac{1}{2}}&=\frac{9}{8}\mathbb{1}_{4}-\frac{1}{2}\left(\hat{\bm{r}}\cdot\bm{S}_{3/2}\right)^2\,,\\
{\cal P}_{\frac{3}{2}\frac{3}{2}}&=-\frac{1}{8}\mathbb{1}_{4}+\frac{1}{2}\left(\hat{\bm{r}}\cdot\bm{S}_{3/2}\right)^2\,,\\
{\cal P}_{10}&=\mathbb{1}_{3}-\left(\hat{\bm{r}}\cdot\bm{S}_{1}\right)^2\,,\\
{\cal P}_{11}&=\left(\hat{\bm{r}}\cdot\bm{S}_{1}\right)^2\,,\\
{\cal P}_{20}&=\mathbb{1}_{5}-\frac{5}{4}\left(\hat{\bm{r}}\cdot\bm{S}_{2}\right)^2+\frac{1}{4}\left(\hat{\bm{r}}\cdot\bm{S}_{2}\right)^4\,,\\
{\cal P}_{21}&=\frac{4}{3}\left(\hat{\bm{r}}\cdot\bm{S}_{2}\right)^2-\frac{1}{3}\left(\hat{\bm{r}}\cdot\bm{S}_{2}\right)^4\,,\\
{\cal P}_{22}&=-\frac{1}{12}\left(\hat{\bm{r}}\cdot\bm{S}_{2}\right)^2+\frac{1}{12}\left(\hat{\bm{r}}\cdot\bm{S}_{2}\right)^4\,.
\end{align}
with $\mathbb{1}_{n}$ an identity matrix in the LDF spin space of dimension $n=2\kappa+1$. The same results can be obtained by taking a specific representation of the spin-$\kappa$ matrices, solving the eigenvalue problem in Eq.~\eqref{eigeq} and using the definition in Eq.~\eqref{defpro}.

Finally, let us note that another useful basis for the operators inside the brackets in Eq.~\eqref{sdp32} depending on the heavy quark pair spin is to use irreducible $O(3)$ tensors
\begin{align}
\bm{{\cal T}}^{ij}_0=&\delta^{ij}=\left({\cal P}^{\rm c.r.}_{10}\right)^{ij}+\left({\cal P}^{\rm c.r.}_{11}\right)^{ij}\,,\\
\bm{{\cal T}}^{ij}_2=&\,\hat{\bm{r}}^i\hat{\bm{r}}^j-\frac{1}{3}\delta^{ij}=\frac{2}{3}\left({\cal P}^{\rm c.r.}_{10}\right)^{ij}-\frac{1}{3}\left({\cal P}^{\rm c.r.}_{11}\right)^{ij}\,.
\end{align}

\section{Matching to NRQCD}
\label{match}

In the following we obtain the static potentials, $V_{\kappa^p\Lambda}^{(0)}$, and the spin-dependent operators in $V^{sa}_{\kappa^p\Lambda\Lambda'}$, $V^{sb}_{\kappa^p\Lambda\Lambda'}$ and $V^{l}_{\kappa^p\Lambda\Lambda'}$ in terms of operator insertions in Wilson loops by expanding NRQCD about the static limit. 

Now, let us define the following strings for the case of hadrons containing a heavy quark-antiquark pair or a heavy quark-quark pair
\begin{align}
\mathcal{O}^{Q\bar{Q}}_{\kappa^p}(t,\,\bm{r},\,\bm{R})=&\chi^{\top}_c(t,\,\bm{x}_2)\phi(t,\,\bm{x}_2,\,\bm{R}){\cal Q}_{Q\bar{Q}\kappa^p}(t,\,\bm{R})\phi(t,\,\bm{R},\,\bm{x}_1)\psi(t,\,\bm{x}_1)\,,\label{stqqb}\\
\mathcal{O}^{QQ}_{\kappa^p}(t,\,\bm{r},\,\bm{R})=&\psi^{\top}(t,\,\bm{x}_2)\phi^{\top}(t,\,\bm{R},\,\bm{x}_2){\cal Q}_{QQ\kappa^p}(t,\,\bm{R})\phi(t,\,\bm{R},\,\bm{x}_1)\psi(t,\,\bm{x}_1)\,,\label{stqq}
\end{align}
with $\psi$ the Pauli spinor fields that annihilate a quark and $\chi$ the one that creates an antiquark, $\chi_c=i\sigma^2\chi^*$. The ${\cal Q}$ operators contain the LDF and are characterized by quantum numbers $\kappa^p$. In Sec.~\ref{light} we provide a list of ${\cal Q}$ operators corresponding to a wide array of heavy exotic states. The Wilson line $\phi$ is defined as follows:
\begin{align}
\phi(t,\bm{x},\bm{y})=P\left\{e^{ig\int_0^1 ds\left(\bm{x}-\bm{y}\right)\cdot \bm{A}(t,\bm{y}+s(\bm{x}-\bm{y}))}\right\}\,,
\end{align}
where $P$ is the path-ordering operator.

The matching condition from NRQCD to the heavy exotic hadron EFT
\begin{align}
\mathcal{O}^h_{\kappa^p}(t,\,\bm{r},\,\bm{R})=\sqrt{Z_{h\kappa^p}}\Psi_{h\kappa^p}(t,\,\bm{r},\,\bm{R})\,,\quad h=Q\bar{Q},\,QQ\,.
\end{align}
The normalization factor is in general a function of $Z_{h\kappa^p}= Z_{h\kappa^p}(\bm{r},\bm{p})$. For simplicity we will not show this dependence explicitly in the rest of the paper. We also suppress the label $h$ except in the steps when the matching procedure for $Q\bar{Q}$ and $QQ$ systems are not the same.

Now, we match the NRQCD and heavy exotic hadron EFT correlators,
\begin{align}
&\langle 0|T\{\mathcal{O}_{\kappa^p}(t/2,\,\bm{r},\,\bm{R})\mathcal{O}^{\dagger}_{\kappa^p}(-t/2,\,\bm{r},\,\bm{R})\}|0\rangle=\sqrt{Z_{\kappa^p}}\langle 0|T\{\Psi_{\kappa^p}(t/2,\,\bm{r},\,\bm{R})\Psi^{\dagger}_{\kappa^p}(-t/2,\,\bm{r},\,\bm{R})\}|0\rangle \sqrt{Z^{\dagger}_{\kappa^p}}\,,\label{corroo}
\end{align}
where we have omitted Dirac deltas on the coordinates on both sides. The right-hand side of Eq.~\eqref{corroo} becomes
\begin{align}
\sqrt{Z_{\kappa^p}}e^{-it h_{\kappa^p}}\sqrt{Z^{\dagger}_{\kappa^p}}=&\sqrt{Z_{\kappa^p}}\frac{1}{t}\int^{t/2}_{-t/2} dt'e^{-i(t/2-t')V_{\kappa^p}^{(0)}(\bm{r})}\left[\mathbb{1}^{Q_1}_{2}\mathbb{1}^{Q_2}_{2}\mathbb{1}^{\rm LDF}_{(2\kappa+1)}\left(1-it\frac{\nabla_r^2}{m_Q}-it\frac{\nabla_R^2}{4m_Q}\right)\right.\nn\\
&\left.-it\frac{1}{m_Q}V_{\kappa^p{\rm SD}}^{(1)}(\bm{r})+\dots\right]e^{-i(t'+t/2)V_{\kappa^p}^{(0)}(\bm{r})}\sqrt{Z^{\dagger}_{\kappa^p}}\label{m1}\,.
\end{align}
The dots stand for ${\rm O}(m^{-1}_Q)$ spin- and velocity-independent potentials and further subleading terms in the heavy quark mass expansion. 

\begin{figure}
\centerline{\includegraphics[width=.7\textwidth]{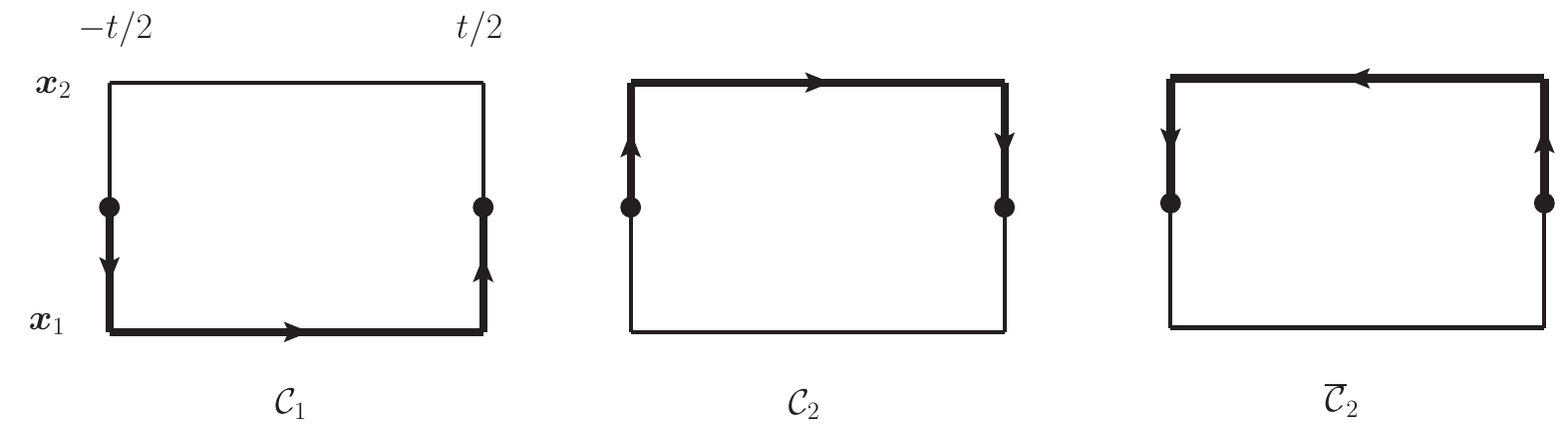}}
\caption{The bold paths correspond to the Wilson line paths ${\cal C}_1$, ${\cal C}_2$ and $\overline{{\cal C}}_2$ associated with each one of the heavy quarks.}
\label{trajectories}
\end{figure}

Let us introduce the following notation:
\begin{align}
&W^{Q\bar{Q}}_{\Box}=P\left\{e^{-ig\int_{{\cal C}_1+\overline{\cal C}_2}dz^{\mu}A^{\mu}(z)}\right\}\,,\\
&W^{QQ}_{\Box}=P\left\{e^{-ig\int_{{\cal C}_1+{\cal C}_2}dz^{\mu}A^{\mu}(z)}\right\}\,,\\
&\langle \dots\rangle^{\kappa^p}_{\Box}=\langle {\cal Q}_{\kappa^p}(t/2,\,\bm{R})\dots {\cal Q}_{\kappa^p}^{\dagger}(-t/2,\,\bm{R})W_{\Box}\rangle\,,
\end{align}
with ${\cal C}_1$, ${\cal C}_2$ and $\overline{{\cal C}}_2$ denoting the upper or lower paths on the rectangular Wilson loop of Fig.~\ref{trajectories}.

Using NRQCD~\cite{Caswell:1985ui,Bodwin:1994jh,Manohar:1997qy}, the right-hand side of Eq.~\eqref{corroo} can be expanded as
\begin{align}
&\langle 1\rangle^{\kappa^p}_{\Box}\mathbb{1}^{Q_1}_{2}\mathbb{1}^{Q_2}_{2}+\dots+i\frac{1}{2m_Q}\int^{t/2}_{-t/2} dt'\left(\langle \bm{D}^2(t',\,\bm{x}_1)\rangle^{\kappa^p}_{\Box }+(-1)^h\langle \bm{D}^2(t',\,\bm{x}_2)\rangle^{\kappa^p}_{\Box }\right)\mathbb{1}^{Q_1}_{2}\mathbb{1}^{Q_2}_{2}\nn\\
&\dots+i\frac{c_F}{2m_Q}\int^{t/2}_{-t/2} dt'\left(\langle \bm{\sigma}_{1}\cdot g\bm{B}(t',\bm{x}_1)\rangle^{\kappa^p}_{\Box }\mathbb{1}^{Q_2}_{2}+\langle \bm{\sigma}_{2}\cdot g\bm{B}(t',\bm{x}_2)\rangle^{\kappa^p}_{\Box }\mathbb{1}^{Q_1}_{2}\right)+\dots\label{m2}
\end{align}
with $(-1)^h=1$ for $h=QQ$ and $(-1)^h=-1$ for $h=Q\bar{Q}$. Comparing Eqs.~\eqref{m1} and \eqref{m2} we obtain 
\begin{align}
\sqrt{Z_{\kappa^p}}e^{-it V_{\kappa^p}^{(0)}(\bm{r})}\sqrt{Z^{\dagger}_{\kappa^p}}=\langle 1\rangle^{\kappa^p}_{\Box}\,.
\end{align}
Now we use the expansion of the static potential in irreducible representations of $D_{\infty h}$ of Eq.~\eqref{edihsp} and arrive at the Wilson loop expressions for the static potentials
\begin{align}
V_{\kappa^p\Lambda}^{(0)}(\bm{r})=&\lim_{t\to\infty}\frac{i}{t}\ln\left({\rm Tr}\left[{\cal P}_{\kappa\Lambda}\langle 1\rangle^{\kappa^p}_{\Box}\right]\right)\,.
\end{align}
Both ${\cal O}(m^{-1}_Q)$ terms in Eq.~\eqref{m2} contribute to heavy quark spin- and angular-momentum-dependent potentials. In the case of the $\bm{D}^2$ operators the contributions come from the third and fourth term of Eq.~\eqref{cdktfd}, which added up are given by Eq.~\eqref{kinlsq}. The potentials match to
\begin{align}
&\sqrt{Z_{\kappa^p}}\frac{1}{t}\int^{t/2}_{-t/2}dt'e^{-i(t/2-t')V_{\kappa^p}^{(0)}(\bm{r})}V_{\kappa^p {\rm SD}}^{(1)}(\bm{r})e^{-i(t'+t/2)V_{\kappa^p}^{(0)}(\bm{r})}\sqrt{Z^{\dagger}_{\kappa^p}}\nn\\
&=-\frac{c_F}{t}\int^{t/2}_{-t/2} dt^{\prime}\bm{S}_{QQ}\cdot \langle g\bm{B}(t^{\prime},\bm{x}_1)\rangle^{\kappa^p}_{\Box}-\int^1_0 ds\,s\,\bm{L}_{QQ}\cdot\langle g\bm{B}(t/2,\bm{z}(s))\rangle^{\kappa^p}_{\Box}\,,\label{is1}
\end{align}
with $\bm{z}(s)=\bm{x}_1+s(\bm{R}-\bm{x}_1)$.

To complete the matching of the heavy quark spin- and angular-momentum-dependent operators in Eq.~\eqref{sdp32}, we decompose the Wilson loop averages in the two LDF spin components using standard tensor decomposition techniques
\begin{align}
\langle \bm{B} \rangle^{\kappa^p}_{\Box}=\sum_{\Lambda\Lambda'}{\cal P}_{\kappa\Lambda}&\left[\delta_{\Lambda\Lambda'}\frac{{\rm Tr}\left[\left(\bm{S}_{\kappa}\cdot{\cal P}^{\rm c.r.}_{10}\right)\cdot\left({\cal P}_{\kappa\Lambda}\langle \bm{B}\rangle^{\kappa^p}_{\Box}{\cal P}_{\kappa\Lambda}\right)\right]}{{\rm Tr}\left[\left(\bm{S}_{\kappa}\cdot{\cal P}^{\rm c.r.}_{10}\right)\cdot\left({\cal P}_{\kappa\Lambda}\bm{S}_{\kappa}{\cal P}_{\kappa\Lambda}\right)\right]}\left({\cal P}^{\rm c.r.}_{10}\cdot\bm{S}_{\kappa}\right)\right.\nn\\
&\left.\quad+\frac{{\rm Tr}\left[\left(\bm{S}_{\kappa}\cdot{\cal P}^{\rm c.r.}_{11}\right)\cdot\left({\cal P}_{\kappa\Lambda}\langle \bm{B}\rangle^{\kappa^p}_{\Box}{\cal P}_{\kappa\Lambda'}\right)\right]}{{\rm Tr}\left[\left(\bm{S}_{\kappa}\cdot{\cal P}^{\rm c.r.}_{11}\right)\cdot\left({\cal P}_{\kappa\Lambda} \bm{S}_\kappa {\cal P}_{\kappa\Lambda'}\right)\right]}\left({\cal P}^{\rm c.r.}_{11}\cdot\bm{S}_{\kappa}\right)\right]{\cal P}_{\kappa\Lambda'}\,.\label{wldec}
\end{align}
Using Eq.~\eqref{wldec} in Eq.~\eqref{is1} we arrive at the following expressions for the heavy quark spin- and angular-momentum-dependent potentials\footnote{We thank Carolin Schlosser for pointing out an error in the way Eqs.~\eqref{pm2} and \eqref{pm3} were displayed in the previous versions of the paper.}:
\begin{align}
V^{sa}_{\kappa^p\Lambda\Lambda'}=&-c_F\lim_{t\to\infty}\frac{\delta_{\Lambda\Lambda'}}{t}\frac{{\rm Tr}\left[{\cal P}_{\kappa\Lambda}\right]}{{\rm Tr}\left[{\cal P}_{\kappa\Lambda}\langle 1 \rangle^{\kappa^p}_{\Box}\right]}\int^{t/2}_{-t/2} dt^{\prime}\frac{{\rm Tr}\left[\left(\bm{S}_{\kappa}\cdot{\cal P}^{\rm c.r.}_{10}\right)\cdot\left({\cal P}_{\kappa\Lambda}\langle \bm{B}(t^{\prime},\bm{x}_1)\rangle^{\kappa^p}_{\Box}{\cal P}_{\kappa\Lambda}\right)\right]}{{\rm Tr}\left[\left(\bm{S}_{\kappa}\cdot{\cal P}^{\rm c.r.}_{10}\right)\cdot\left({\cal P}_{\kappa\Lambda}\bm{S}_{\kappa}{\cal P}_{\kappa\Lambda}\right)\right]}\,,\label{pm1}\\
V^{sb}_{\kappa^p\Lambda\Lambda'}=&-\frac{c_F}{2}\lim_{t\to\infty}\sqrt{\frac{{\rm Tr}\left[{\cal P}_{\kappa\Lambda}\right]{\rm Tr}\left[{\cal P}_{\kappa\Lambda'}\right]}{{\rm Tr}\left[{\cal P}_{\kappa\Lambda}\langle 1 \rangle^{\kappa^p}_{\Box}\right]{\rm Tr}\left[{\cal P}_{\kappa\Lambda'}\langle 1 \rangle^{\kappa^p}_{\Box}\right]}}\frac{V^{(0)}_{\kappa^p\Lambda}-V^{(0)}_{\kappa^p\Lambda'}}{\sin\left[\left(V^{(0)}_{\kappa^p\Lambda}-V^{(0)}_{\kappa^p\Lambda'}\right)\frac{t}{2}\right]}\nn\\
&\int^{t/2}_{-t/2} dt^{\prime}\frac{{\rm Tr}\left[\left(\bm{S}_{\kappa}\cdot{\cal P}^{\rm c.r.}_{11}\right)\cdot\left({\cal P}_{\kappa\Lambda}\langle g\bm{B}(t^{\prime},\bm{x}_1)\rangle^{\kappa^p}_{\Box}{\cal P}_{\kappa\Lambda'}\right)\right]}{{\rm Tr}\left[\left(\bm{S}_{\kappa}\cdot{\cal P}^{\rm c.r.}_{11}\right)\cdot\left({\cal P}_{\kappa\Lambda} \bm{S}_\kappa {\cal P}_{\kappa\Lambda'}\right)\right]}\,,\label{pm2}\\
V^{l}_{\kappa^p\Lambda\Lambda'}=&-\lim_{t\to\infty}\sqrt{\frac{{\rm Tr}\left[{\cal P}_{\kappa\Lambda}\right]{\rm Tr}\left[{\cal P}_{\kappa\Lambda'}\right]}{{\rm Tr}\left[{\cal P}_{\kappa\Lambda}\langle 1 \rangle^{\kappa^p}_{\Box}\right]{\rm Tr}\left[{\cal P}_{\kappa\Lambda'}\langle 1 \rangle^{\kappa^p}_{\Box}\right]}}\frac{\left(V^{(0)}_{\kappa^p\Lambda}-V^{(0)}_{\kappa^p\Lambda'}\right)\frac{t}{2}}{\sin\left[\left(V^{(0)}_{\kappa^p\Lambda}-V^{(0)}_{\kappa^p\Lambda'}\right)\frac{t}{2}\right]}\nn\\
&\int^1_0 ds\,s \frac{{\rm Tr}\left[\left(\bm{S}_{\kappa}\cdot{\cal P}^{\rm c.r.}_{11}\right)\cdot\left({\cal P}_{\kappa\Lambda}\langle g\bm{B}(t/2,\bm{z}(s))\rangle^{\kappa^p}_{\Box}{\cal P}_{\kappa\Lambda'}\right)\right]}{{\rm Tr}\left[\left(\bm{S}_{\kappa}\cdot{\cal P}^{\rm c.r.}_{11}\right)\cdot\left({\cal P}_{\kappa\Lambda} \bm{S}_\kappa {\cal P}_{\kappa\Lambda'}\right)\right]}\,\,.\label{pm3}
\end{align}
The computation of the traces not involving $\bm{B}$ can be found in Appendix~\ref{tracecomp}. In Appendix~\ref{shortcomp} we compare with known results for the matching of the hybrid quarkonium static potentials and the heavy-quark-spin-dependent potentials in the short-distance limit.

\section{Light-quark and gluon operators}
\label{light}

In this section we provide with specific LDF operators for Eqs.~\eqref{stqqb} and \eqref{stqq} interpolating for a wide array of possible heavy exotic hadrons. Let us first introduce the notation. The light quark fields are standard Dirac fermions represented by $q_{f\alpha}^a(t,\bm{x})$ where $a$ is the color index and $\alpha$ the spin index, and $f$ is the isospin index. The ${\cal C}^{j_3\,m_3}_{j_1\,m_1\,j_2\,m_2}$ is a Clebsch-Gordan coefficient, the projector $P_+=(1+\gamma^0)/2$ and the polarization vectors $\bm{e}_{+1}=-(1,\,i,\,0)/\sqrt{2}$, $\bm{e}_{-1}=(1,\,-i,\,0)/\sqrt{2}$, $\bm{e}_{0}=(0,\,0,\,1)$. $T^a$ is the standard fundamental representation of the generators of $SU(3)$. We also use the $\bar{3}$ and $6$ tensor invariants from Ref.~\cite{Brambilla:2005yk}
\begin{align}
&\underline{T}^l_{ij}  = \frac{1}{\sqrt{2}} \epsilon_{lij},\quad i,\,j,\,l=1,2,3\,
\label{Tabg}
\\
& \nn\\
& \underline{\Sigma}^\sigma_{ij}\quad i,\,j=1,2,3\,\quad \sigma=1,..,6\nn\\
&\underline{\Sigma}^1_{11}  = \underline{\Sigma}^4_{22} = \underline{\Sigma}^6_{33} = 1,
\nn\\
&\underline{\Sigma}^2_{12}  = \underline{\Sigma}^2_{21} = 
 \underline{\Sigma}^3_{13}  = \underline{\Sigma}^3_{31} = 
 \underline{\Sigma}^5_{23}  = \underline{\Sigma}^5_{32} = \frac{1}{\sqrt{2}}, 
\end{align}
all other entries are zero. Both $\underline{T}^l_{ij}$ and $\underline{\Sigma}^\sigma_{ij}$ are real; $\underline{T}^l_{ij}$ is totally antisymmetric and $\underline{\Sigma}^\sigma_{ij}$  symmetric in the $i$ and $j$ indices.
They satisfy the orthogonality and normalization relations:
\begin{align}
\sum_{ij=1}^3 \underline{T}^{l_1}_{ij} \, \underline{T}^{l_2}_{ij} =
\delta^{l_1l_2}\,, \qquad
\sum_{ij=1}^3 \underline{\Sigma}^{\sigma_1}_{ij} \, \underline{\Sigma}^{\sigma_2}_{ij} = 
\delta^{\sigma_1\sigma_2}\,, \qquad
\sum_{ij=1}^3 \underline{T}^l_{ij} \, \underline{\Sigma}^\sigma_{ij} = 0\,.
\end{align}

In the case of a heavy quark-antiquark pair, these can be in singlet or octet states. The first case corresponds to the standard quarkonium. For the latter we can construct the following operators interpolating for hybrid quarkonium
\begin{align}
{\cal Q}^{\alpha}_{1^{+-}}(t,\bm{x})&= \left(\bm{e}^{\dagger}_{\alpha}\cdot\bm{B}(t,\bm{x})\right)\,,\label{Qhybrid}\\
{\cal Q}^{\alpha}_{1^{--}}(t,\bm{x})&= \left(\bm{e}^{\dagger}_{\alpha}\cdot\bm{E}(t,\bm{x})\right)\,,\\
{\cal Q}^{\alpha}_{2^{+-}}(t,\bm{x})&=  {\cal C}^{2\,\alpha}_{1\,m_1\,1\,m_2}\left(\bm{e}^{\dagger}_{m_1}\cdot\bm{D}(t,\bm{x})\right)\left(\bm{e}^{\dagger}_{m_2}\cdot\bm{E}(t,\bm{x})\right)\,,\\
{\cal Q}^{\alpha}_{2^{--}}(t,\bm{x})&=  {\cal C}^{2\,\alpha}_{1\,m_1\,1\,m_2}\left(\bm{e}^{\dagger}_{m_1}\cdot\bm{D}(t,\bm{x})\right)\left(\bm{e}^{\dagger}_{m_2}\cdot\bm{B}(t,\bm{x})\right)\,.
\end{align}
Light-quark operators interpolating for isospin $I=0$ tetraquark states can be constructed as
\begin{align}
{\cal Q}_{0^{++}}(t,\bm{x})&= \left[\bar{q}(t,\bm{x})T^a q(t,\bm{x})\right]T^a\,,\label{tqo1}\\
{\cal Q}_{0^{-+}}(t,\bm{x})&= \left[\bar{q}(t,\bm{x})\gamma^5 T^a q(t,\bm{x})\right]T^a\,,\\
{\cal Q}^{\alpha}_{1^{++}}(t,\bm{x})&= \left[\bar{q}(t,\bm{x})\left(\bm{e}^{\dagger}_{\alpha}\cdot\bm{\gamma}\right)\gamma^5 T^a q(t,\bm{x})\right]T^a\,,\label{tqo2}\\
{\cal Q}^{\alpha}_{1^{--}}(t,\bm{x})&= \left[\bar{q}(t,\bm{x})\left(\bm{e}^{\dagger}_{\alpha}\cdot\bm{\gamma}\right)T^a q(t,\bm{x})\right]T^a\,,\label{tqo3}\\
{\cal Q}^{\alpha}_{1^{+-}}(t,\bm{x})&= \left[\bar{q}(t,\bm{x})\left(\bm{e}^{\dagger}_{\alpha}\cdot\left(\bm{\gamma}\times \bm{\gamma}\right)\right)T^a q(t,\bm{x})\right]T^a\,,\label{tqo4}\\
{\cal Q}^{\alpha}_{2^{+-}}(t,\bm{x})&= {\cal C}^{2\,\alpha}_{1\,m_1\,1\,m_2}\left[\bar{q}(t,\bm{x})\left(\bm{e}^{\dagger}_{m_1}\cdot\bm{D}(t,\,\bm{x})\right)\left(\bm{e}^{\dagger}_{m_2}\cdot\bm{\gamma}\right)T^a q(t,\bm{x})\right]T^a\,.\label{tqo5}
\end{align}
The isospin $I=1$ tetraquark states can be interpolated by similar operators by just adding $\bm{e}_{I_3}\cdot\bm{\tau}$ in between the light quark fields. Let us note that for the states with $\kappa=0$ the operators in Eq.~\eqref{sdp32} vanish.

Quarkonium-pentaquark states with $I=1/2$ and $I_3=\pm1/2$ are interpolated by the operators read as
\begin{align}
{\cal Q}^{\alpha}_{I_3(1/2)^{+}}(t,\bm{x})=& \left(\delta_{\alpha\beta_1}\sigma^2_{\beta_2\beta_3}+\delta_{\alpha\beta_2}\sigma^2_{\beta_1\beta_3}+\delta_{\alpha\beta_3}\sigma^2_{\beta_1\beta_2}\right)\left(\delta_{I_3f_1}\tau^2_{f_2f_3}+\delta_{I_3f_2}\tau^2_{f_3f_1}+\delta_{I_3f_3}\tau^2_{f_1f_2}\right)\left(\underline{T}^{k}_{l_1l_2}T_{kl_3}^a\right.\nn\\
&\left.+\underline{T}^{k}_{l_1l_3}T_{kl_2}^a+\underline{T}^{k}_{l_2l_3}T_{kl_1}^a\right)\left(P_+q_{l_1f_1}(t,\bm{x})\right)^{\beta_1} \left(P_+q_{l_2f_2}(t,\bm{x})\right)^{\beta_2} \left(P_+q_{l_3f_3}(t,\bm{x})\right)^{\beta_3}T^a\,.
\end{align}

Operators interpolating for double heavy baryons with the heavy quarks in a $\bar{3}$ state
\begin{align}
{\cal Q}^{\alpha}_{(1/2)^+}(t,\bm{x})&= \underline{T}^l \left[P_+q^{l}(t,\bm{x})\right]^\alpha\,,\\
{\cal Q}^{\alpha}_{(1/2)^-}(t,\bm{x})&= \underline{T}^l\left[P_+\gamma^5q^l(t,\bm{x})\right]^\alpha\,,\\
{\cal Q}^{\beta}_{(3/2)^-}(t,\bm{x})&= {\cal C}^{3/2\,\beta}_{1\,m\,1/2\,\alpha}\underline{T}^l\left[\left(\bm{e}^{\dagger}_{m}\cdot\bm{D}\right) \left(P_+q(t,\bm{x})\right)^\alpha\right]^l\,,\\
{\cal Q}^{\beta}_{(3/2)^+}(t,\bm{x})&= {\cal C}^{3/2\,\beta}_{1\,m\,1/2\,\alpha}\underline{T}^l\left[\left(\bm{e}^{\dagger}_{m}\cdot\bm{D}\right)\left(P_+\gamma^5q(t,\bm{x})\right)^\alpha\right]^l\,,
\end{align}
if the heavy quarks are in a $6$ color state, gluonic fields need to be added
\begin{align}
{\cal Q}^{\beta}_{(3/2)^-}(t,\bm{x})&= {\cal C}^{3/2\,\beta}_{1\,m\,1/2\,\alpha}\underline{\Sigma}^\sigma {\rm Tr}\left[\left(\bm{e}^{\dagger}_{m}\cdot\bm{E}\right)\underline{\Sigma}^\sigma\underline{T}^l\right]\left(P_+q^l(t,\bm{x})\right)^\alpha\,,\\
{\cal Q}^{\beta}_{(3/2)^+}(t,\bm{x})&= {\cal C}^{3/2\,\beta}_{1\,m\,1/2\,\alpha}\underline{\Sigma}^\sigma {\rm Tr}\left[\left(\bm{e}^{\dagger}_{m}\cdot\bm{B}\right)\underline{\Sigma}^\sigma\underline{T}^l\right]\left(P_+q^l(t,\bm{x})\right)^\alpha\end{align}

Finally, one can also construct open heavy flavor tetraquarks in a similar manner to Eqs.~\eqref{tqo1}-\eqref{tqo5}. 
\begin{align}
{\cal Q}_{0^{-}}(t,\bm{x})&= \left[\bar{q}(t,\bm{x})\underline{T}^l \gamma^2q^*(t,\bm{x})\right]\underline{T}^l\,,\label{fftqo1}\\
{\cal Q}_{0^{+}}(t,\bm{x})&= \left[\bar{q}(t,\bm{x})\gamma^5 \underline{T}^l \gamma^2q^*(t,\bm{x})\right]\underline{T}^l\,,\label{fftqo2}\\
{\cal Q}^{\alpha}_{1^{-}}(t,\bm{x})&= \left[\bar{q}(t,\bm{x})\left(\bm{e}^{\dagger}_{\alpha}\cdot\bm{\gamma}\right)\underline{T}^l\gamma^5\gamma^2q^*(t,\bm{x})\right]\underline{T}^l\,,\label{fftqo3}\\
{\cal Q}^{\alpha}_{1^{+}}(t,\bm{x})&= \left[\bar{q}(t,\bm{x})\left(\bm{e}^{\dagger}_{\alpha}\cdot\bm{\gamma}\right)\underline{T}^l \gamma^2q^*(t,\bm{x})\right]\underline{T}^l\,,\label{fftqo4}\\
{\cal Q}^{\alpha}_{2^{-}}(t,\bm{x})&= {\cal C}^{2\,\alpha}_{1\,m_1\,1\,m_2}\left[\bar{q}(t,\bm{x})\left(\bm{e}^{\dagger}_{m_1}\cdot\bm{D}\right)\left(\bm{e}^{\dagger}_{m_2}\cdot\bm{\gamma}\right)\underline{T}^l \gamma^2q^*(t,\bm{x})\right]\underline{T}^l\,.\label{fftqo6}
\end{align}
The isospin $I=1$ open heavy flavor tetraquark states can be interpolated by similar operators by adding $\bm{e}_{I_3}\cdot\left(\tau^2\bm{\tau}\right)$ in between the light quark fields.

\section{Conclusions}

We have put forward a general formalism to build EFTs for exotic hadrons with two heavy quarks as well as double heavy baryons. It is based on NRQCD \cite{Caswell:1985ui,Thacker:1990bm} and the strong coupling regime of pNRQCD \cite{Brambilla:1999xf}. Hence, the basic assumptions are that (i) the heavy quark masses are larger than any other energy scale in the system, and (ii) the heavy quark binding energies are smaller than the typical hadronic scale $\lQ$. We pose no extra assumption on the relative size of the typical momentum exchanges between the heavy quarks $\sim 1/r$ and $\lQ$ such as in previous works~\cite{Brambilla:2005yk,Mehen:2019cxn,Brambilla:2018pyn,An:2018cln,Brambilla:2019jfi}. The effective theory consists of wave function fields interacting with potentials, which are a function of $r$ and $\lQ$. At leading order in the $1/m_Q$ expansion, these potentials correspond to the Born-Oppenheimer approximation. They are heavy quark spin- and angular-momentum-independent, but, in general, they depend on the spin of the LDF. At next-to-leading order in the $1/m_Q$ expansion, heavy-quark spin- and angular-momentum-dependent potentials arise. We provide formulas to calculate, both leading-order and next-to-leading order potentials, from QCD for any spin of the LDF in terms of operator insertions in the Wilson loop. These are suitable to be computed in lattice QCD with only light quarks and gluons. 

It is important to emphasize that unlike heavy quarkonium, in general, for exotic hadrons with $\kappa\neq 0$ the heavy quark spin- and orbital-angular-momentum (velocity) effects start at order $1/m_Q$, and not at order $1/m_Q^2$. This is important since a heavy quark spin dependence analogous to the heavy quarkonium one is often assumed in models, see for example Refs.~\cite{Barnes:1981ac,Cornwall:1982zn,Lebed:2017xih}. We hope that our general formalism makes this point clear.

For a given spin representation of the $O(3)$ symmetry group ($\kappa^p$) of the LDF, we have seen that $1/m_Q$ terms mix different representations of the $D_{\infty h}$ group. This was already noted in the specific case of quarkonium hybrids in Refs.~\cite{Soto:2017one,Brambilla:2018pyn,Brambilla:2019jfi}. Furthermore, the $1/m_Q$ terms may also mix different representations of the $O(3)$ group. This was already noted, again in the specific case of quarkonium hybrids, in Refs.~\cite{Oncala:2017hop,Soto:2017one}. Indeed, the operators in the Lagrangian of Eq.~\eqref{sdp32} can be generalized to consider off-diagonal terms with $\kappa=\kappa'\pm1$ and $p=p'$, that will produce heavy exotic hadrons with mixed $\bm{s}_{QQ}$ and $l$. Moreover, the terms in Eq.~\eqref{lsqntr} that do not contribute to the Lagrangian of Eq.~\eqref{sdp32} could also produce off-diagonal operators in $\kappa$ and other quantum numbers. A full analysis of these mixing operators and its matching is left for future work. These mixing terms become relevant for the spectrum at order $1/m_Q^2$, although their contribution is suppressed by the energy gap between $\kappa^p$ representations. However it may become important if there are accidental degeneracies in the spectrum of both representations~\cite{Oncala:2017hop}. 

The EFT built in this way is expected to be reliable for the ground and excited states with binding energies $E\ll \lQ$ about the ground state. When the binding energies approach the hadronic scale, both light hadron resonances and hadron pairs with a single heavy quark each should somehow be included in the approach. Since they are not, at this scale our EFT becomes a model, which, however, fully incorporates the NRQCD symmetries, in particular the heavy quark spin-symmetry breaking pattern. It may also happen, and in fact it often does, that the ground state of the exotic hadron itself is close to a hadron pair threshold. In that case, again, the hadrons forming the threshold should be included in the EFT. String breaking data of lattice QCD \cite{Bali:2005fu,Bulava:2019iut} suggest that for heavy-light hadron pair thresholds, threshold effects are only noticeable in a tiny energy band around the threshold of a few tens of MeV. In case that the ground and excited states are away from these thresholds by more than a few tens of MeV, our EFT becomes again a reasonable model. Unfortunately, there is no such information for thresholds involving light hadrons.

Since pions have masses parametrically smaller than $\lQ$, they could be incorporated to the EFT in a model-independent way, by using a formalism similar to Heavy Baryon Chiral Perturbation Theory~\cite{Burdman:1992gh}, that was used in Refs.~\cite{TarrusCastella:2019rit,Pineda:2019mhw} for hybrids and tetraquarks. The outcome would be similar to hadronic chiral theories, see for instance~\cite{casalbuoni:1996pg}, but with $r$-dependent low-energy constants. Soft photons could also be incorporated by matching NRQCD coupled to e.m. to the effective theory coupled to soft photons.

Finally, we have listed a number of light quark and gluon operators at the NRQCD level that describe most of the exotic hadrons containing two heavy quarks that are being considered nowadays as well as double heavy baryons. In an accompanying paper~\cite{Soto:2020pfa}, we present a concrete application to doubly heavy baryons.
\label{concl}

\section*{Acknowledgements}

J.S. acknowledges financial support from  the 2017-SGR-929 grant from the Generalitat de Catalunya and the FPA2016-76005-C2-1-P and FPA2016-81114-P projects from Ministerio de Ciencia, Innovaci\'on y Universidades. J.T.C. acknowledges financial support from the European Union's Horizon 2020 research and innovation programme under the Marie Sk\l{}odowska--Curie Grant Agreement No. 665919. He has also been supported in part by the Spanish Grants No. FPA2017-86989-P and No. SEV-2016-0588 from the Ministerio de Ciencia, Innovaci\'on y Universidades, and the Grant No. 2017-SGR-1069 from the Generalitat de Catalunya. This research was supported by the Munich Institute for Astro- and Particle Physics (MIAPP) which is funded by the Deutsche Forschungsgemeinschaft (DFG, German Research Foundation) under Germany's Excellence Strategy – EXC-2094 – 390783311.

\appendix

\section{Simplification of the Wilson loop with a kinetic operator insertion}\label{wlsimp}

The $1/m_Q$ contribution to the left-hand side of Eq.~\eqref{corroo} produced by the kinetic operator is
\begin{align}
&\int^{t/2}_{-t/2} dt'\langle \bm{D}^2(t',\,\bm{x}_1)\rangle^{\kappa^p}_{\Box}=\nn\\
&\int^{t/2}_{-t/2}dt'\langle 0 |\dots \phi(t/2,\bm{R},\bm{x}_1)\phi(t/2,t',\bm{x}_1)\bm{D}^2(t',\bm{x}_1)\phi(t',-t/2,\bm{x}_1)\phi(-t/2,\bm{x}_1,\bm{R})\dots|0\rangle\label{lsq1}
\end{align}
where the dots stand for the terms that we do not display explicitly, namely the LDF operators and Wilson lines that do not explicitly depend on $\bm{x}_1$. These terms will be unaffected by the  manipulations that we will perform in this appendix. We also omit the label $h=Q\bar{Q},\,QQ$ and comment about the differences between these two cases at the end of the appendix.

The Wilson lines are defined as follows:
\begin{align}
&\phi(t',t,\bm{x})=P\left\{e^{-ig\int_t^{t'}dt''A_0(t'',\,\bm{x})}\right\}\,,\label{stwl} \\ 
&\phi(t,\bm{x},\bm{y})=P\left\{e^{ig\int_0^1 ds (\bm{x}-\bm{y})\cdot\bm{A}(t,\bm{y}+s(\bm{x}-\bm{y}))}\right\}\,,
\end{align}
where $P$ stands for path ordered. We will use the following equalities~\cite{Eichten:1980mw,Brambilla:2000gk}
\begin{align}
&\phi(t'',t',\bm{x})\phi(t',t,\bm{x})=\phi(t'',t,\bm{x})\,,\label{wleq1}\\
&\bm{D}(\bm{x},t')\phi(t',t,\bm{x})=\phi(t',t,\bm{x})\bm{D}(\bm{x},t)+i{\cal O}_E(t',t,\bm{x})\,,\label{wleq2}\\
&\bm{D}(\bm{x},t)\phi(t,\bm{x},\bm{y})=\phi(t,\bm{x},\bm{y})\bm{\nabla}_{x}+i{\cal O}_B(t,\bm{x},\bm{y})\,,\label{wleq3}
\end{align}
with the abbreviated notation for the following strings
\begin{align}
&{\cal O}_E(t',t,\bm{x})=\int^{t'}_t dt''\phi(t',t'',\bm{x})g\bm{E}(t'',\bm{x})\phi(t'',t,\bm{x})\,,\\
&{\cal O}_B(t,\bm{x},\bm{y})=\int^1_0 ds\,s\,\phi(t,\bm{x},\bm{z}(s))\left(\bm{x}-\bm{y}\right)\times g\bm{B}(t,\bm{z}(s))\phi(t,\bm{z}(s),\bm{y})\,,\label{defob}
\end{align}
with $\bm{z}(s)=\bm{y}+s(\bm{x}-\bm{y})$. Using Eqs.~\eqref{wleq1}-\eqref{wleq3} in Eq.~\eqref{lsq1} one can arrive at
\begin{align}
\langle \bm{D}^2(t',\,\bm{x}_1)\rangle^{\kappa^p}_{\Box}=&\frac{\bm{\nabla}_{x_1}}{2}\cdot\left(\langle 1\rangle^{\kappa^p}_{\Box}\bm{\nabla}_{x_1}+2i\langle {\cal O}_B(-t/2,\bm{x}_1,\bm{R}) \rangle^{\kappa^p}_{\Box}+2i\langle {\cal O}_E(t',-t/2,\bm{x}_1) \rangle^{\kappa^p}_{\Box}\right)+\left(\bm{\nabla}_{x_1}\langle 1\rangle^{\kappa^p}_{\Box}\right.\nn\\
&\left.+2i\langle {\cal O}_B(t/2,\bm{R},\bm{x}_1) \rangle^{\kappa^p}_{\Box}-2i\langle {\cal O}_E(t/2,t',\bm{x}_1) \rangle^{\kappa^p}_{\Box}\right)\cdot\frac{\bm{\nabla}_{x_1}}{2}-\langle {\cal O}_B(t/2,\bm{R},\bm{x}_1){\cal O}_E(t',-t/2,\bm{x}_1)  \rangle^{\kappa^p}_{\Box}\nn\\
&+\langle {\cal O}_E(t/2,t',\bm{x}_1){\cal O}_B(-t/2,\bm{x}_1,\bm{R})  \rangle^{\kappa^p}_{\Box}+\langle {\cal O}_E(t/2,t',\bm{x}_1){\cal O}_E(t',-t/2,\bm{x}_1)  \rangle^{\kappa^p}_{\Box}\nn\\
&+\langle {\cal O}_B(t/2,\bm{R},\bm{x}_1){\cal O}_B(-t/2,\bm{x}_1,\bm{R})  \rangle^{\kappa^p}_{\Box}\,,\label{lsq5}
\end{align}
where the strings in the Wilson loops are understood as replacing that corresponding segment of the loop. 

Now it is convenient to work out the following identity
\begin{align}
&\bm{\nabla}_{x_1}\langle 1\rangle^{\kappa^p}_{\Box}=-i\langle {\cal O}_B(t/2,\bm{R},\bm{x}_1)\rangle^{\kappa^p}_{\Box}+i\langle {\cal O}_E(t/2,-t/2,\bm{x}_1)\rangle^{\kappa^p}_{\Box}+i\langle {\cal O}_B(-t/2,\bm{x}_1,\bm{R})\rangle^{\kappa^p}_{\Box}+\langle 1\rangle^{\kappa^p}_{\Box}\bm{\nabla}_{x_1}\,,\label{lsq2}
\end{align}
which allows one to write Eq.~(\ref{lsq5}) as
\begin{align}
\langle \bm{D}^2(t',\,\bm{x}_1)\rangle^{\kappa^p}_{\Box}=&\frac{\bm{\nabla}_{x_1}^2}{2}\langle 1\rangle^{\kappa^p}_{\Box}+
\langle 1\rangle^{\kappa^p}_{\Box}\frac{\bm{\nabla}_{x_1}^2}{2} \nn\\
+&\frac{\bm{\nabla}_{x_1}}{2}\cdot\left(i\langle {\cal O}_B(-t/2,\bm{x}_1,\bm{R}) \rangle^{\kappa^p}_{\Box}+i\langle {\cal O}_B(t/2,\bm{R},\bm{x}_1) \rangle^{\kappa^p}_{\Box}-i\langle {\cal O}_E(t/2,t',\bm{x}_1) \rangle^{\kappa^p}_{\Box}+
i\langle {\cal O}_E(t',-t/2,\bm{x}_1) \rangle^{\kappa^p}_{\Box}\right)\nn\\
&\left(i\langle {\cal O}_B(t/2,\bm{R},\bm{x}_1) \rangle^{\kappa^p}_{\Box}+i\langle {\cal O}_B(-t/2,\bm{x}_1,\bm{R}) \rangle^{\kappa^p}_{\Box}-i\langle {\cal O}_E(t/2,t',\bm{x}_1) \rangle^{\kappa^p}_{\Box}+i\langle {\cal O}_E(t',-t/2,\bm{x}_1) \rangle^{\kappa^p}_{\Box}\right)\cdot\frac{\bm{\nabla}_{x_1}}{2}\nn\\
&-\langle {\cal O}_B(t/2,\bm{R},\bm{x}_1){\cal O}_E(t',-t/2,\bm{x}_1)  \rangle^{\kappa^p}_{\Box}
+\langle {\cal O}_E(t/2,t',\bm{x}_1){\cal O}_B(-t/2,\bm{x}_1,\bm{R})  \rangle^{\kappa^p}_{\Box}\nn\\
&+\langle {\cal O}_E(t/2,t',\bm{x}_1){\cal O}_E(t',-t/2,\bm{x}_1)  \rangle^{\kappa^p}_{\Box}
+\langle {\cal O}_B(t/2,\bm{R},\bm{x}_1){\cal O}_B(-t/2,\bm{x}_1,\bm{R})  \rangle^{\kappa^p}_{\Box}\,\label{lsqntr}
\end{align}
Note also that due to time reversal the identities
\begin{align}
&\int^{t/2}_{-t/2} dt' \langle {\cal O}_E(t',-t/2,\bm{x}_1)\rangle^{\kappa^p}_{\Box}=\int^{t/2}_{-t/2} dt' \langle {\cal O}_E(t/2,t',\bm{x}_1)\rangle^{\kappa^p}_{\Box}=\frac{1}{2}\int^{t/2}_{-t/2} dt' \langle {\cal O}_E(t/2,-t/2,\bm{x}_1)\rangle^{\kappa^p}_{\Box}\label{lsq3}\,,
\end{align}
and
\begin{align}
&\langle {\cal O}_B(t/2,\bm{R},\bm{x}_1) \rangle^{\kappa^p}_{\Box}=\langle {\cal O}_B(-t/2,\bm{x}_1,\bm{R}) \rangle^{\kappa^p}_{\Box}\,,\label{lsq4}
\end{align}
are fulfilled.
Using Eqs.~\eqref{lsq3}-\eqref{lsq4} into Eq.~\eqref{lsqntr} we arrive at
\begin{align}
\langle\bm{D}^2(t',\,\bm{x}_1)\rangle^{\kappa^p}_{\Box}=&\frac{1}{2}\bm{\nabla}^2_{x_1}\langle 1\rangle^{\kappa^p}_{\Box}+\frac{1}{2}\langle 1\rangle^{\kappa^p}_{\Box}\bm{\nabla}^2_{x_1}+i\bm{\nabla}_{x_1} \cdot\langle {\cal O}_B(t/2,\bm{R},\bm{x}_1) \rangle^{\kappa^p}_{\Box}+i\langle {\cal O}_B(t/2,\bm{R},\bm{x}_1) \rangle^{\kappa^p}_{\Box}\cdot\bm{\nabla}_{x_1}\nn\\
&-\langle {\cal O}_B(t/2,\bm{R},\bm{x}_1){\cal O}_E(t',-t/2,\bm{x}_1)  \rangle^{\kappa^p}_{\Box}+\langle {\cal O}_E(t/2,t',\bm{x}_1){\cal O}_B(-t/2,\bm{x}_1,\bm{R})  \rangle^{\kappa^p}_{\Box}\nn\\
&+\langle {\cal O}_E(t/2,t',\bm{x}_1){\cal O}_E(t',-t/2,\bm{x}_1)  \rangle^{\kappa^p}_{\Box}+\langle {\cal O}_B(t/2,\bm{R},\bm{x}_1) {\cal O}_B(-t/2,\bm{x}_1,\bm{R})  \rangle^{\kappa^p}_{\Box}\,.\label{cdktfd}
\end{align}
The term that matches to the $\bm{L}_{QQ}\cdot \bm{S}_\kappa$ operator is
\begin{align}
i\bm{\nabla}_{x_1} \cdot\langle {\cal O}_B(t/2,\bm{R},\bm{x}_1) \rangle^{\kappa^p}_{\Box}+
i\langle {\cal O}_B(t/2,\,\bm{R},\,\bm{x}_1) \rangle^{\kappa^p}_{\Box}\cdot\bm{\nabla}_{x_1}=&-i\bm{r}^i \epsilon_{ijk}\int^1_0 ds\,s\,\langle \bm{B}^j(t/2,\bm{z}(s))\rangle^{\kappa^p}_{\Box}\left(\frac{1}{2}\bm{\nabla}_{R}+\bm{\nabla}_{r}\right)^k+\dots\nn\\
=&\int^1_0 ds\,s\,\langle \bm{B}^j(t/2,\bm{z}(s))\rangle^{\kappa^p}_{\Box}\bm{L}_{QQ}^j+\dots
\end{align}
In the heavy quark pair case the contributions from $\bm{D}^2(t,\bm{x}_2)$ are analogous except by two relative a minus signs that compensate each other,
\begin{align}
i\bm{\nabla}_{x_2} \cdot\langle {\cal O}_B(t/2,\bm{R},\bm{x}_2) \rangle^{\kappa^p}_{\Box}+i\langle {\cal O}_B(t/2,\,\bm{R},\,\bm{x}_2) \rangle^{\kappa^p}_{\Box}\cdot\bm{\nabla}_{x_2}=&\,i\bm{r}^i \epsilon_{ijk}\int^1_0 ds\,s\,\langle \bm{B}^j(t/2,\bm{z}'(s))\rangle^{\kappa^p}_{\Box}\left(\frac{1}{2}\bm{\nabla}_{R}-\bm{\nabla}_{r}\right)^k+\dots\nn\\
=\int^1_0 ds\,s\,\langle \bm{B}^j(t/2,\bm{z}'(s))\rangle^{\kappa^p}_{\Box}\bm{L}_{QQ}^j+\dots=&\int^1_0 ds\,s\,\langle \bm{B}^j(t/2,\bm{z}(s))\rangle^{\kappa^p}_{\Box}\bm{L}_{QQ}^j+\dots
\end{align}
with $\bm{z}'(s)=\bm{R}+s(\bm{x}_2-\bm{R})$. Adding up the two contributions we arrive at
\begin{align}
&\int^{t/2}_{-t/2}dt'\left(\langle\bm{D}^2(t',\bm{x}_1)\rangle^{\kappa^p}_{\Box}+\langle\bm{D}^2(t',\bm{x}_2)\rangle^{\kappa^p}_{\Box}\right)=2t\int^1_0 ds\,s\,\langle\bm{B}(t/2,\bm{z}(s))\rangle^{\kappa^p}_{\Box}\cdot\bm{L}_{QQ}+\dots\label{kinlsq}
\end{align}
In the case of a heavy quark-antiquark pair the kinetic operator of the heavy antiquark has a relative minus sign with respect to the heavy quark one. However, the path of the Wilson line of the heavy antiquark goes in the opposite direction to the one of the heavy quark, see Fig.~\ref{trajectories}. This opposite trajectory for the antiquark generates an additional minus sign in Eq.~\eqref{defob} since that string depends on the initial and end points. The two minus signs compensate each other and the result for a heavy quarks-antiquark pair is also the one given by Eq.~\eqref{kinlsq}.

\section{Traces of projectors and spin matrices}\label{tracecomp}

The trace of the projectors is
\begin{align}
{\rm Tr}\left[{\cal P}_{\kappa\Lambda}\right]&=2-\delta_{\Lambda 0}\,.
\end{align}
The traces of projectors and spin matrices in Eqs.~\eqref{pm1}-\eqref{pm3} can be expressed as sums of Clebsch-Gordan coefficients
\begin{align}
{\rm Tr}\left[\left(\bm{S}_{\kappa}\cdot{\cal P}^{\rm c.r.}_{10}\right)\cdot\left({\cal P}_{\kappa\Lambda} \bm{S}_\kappa {\cal P}_{\kappa\Lambda'}\right)\right]&=2\Lambda^2\delta_{\Lambda\Lambda'}\,,\\
{\rm Tr}\left[\left(\bm{S}_{\kappa}\cdot{\cal P}^{\rm c.r.}_{11}\right)\cdot\left({\cal P}_{\kappa\Lambda} \bm{S}_\kappa {\cal P}_{\kappa\Lambda'}\right)\right]&=\left(\kappa+\Lambda\right)\left(\kappa-\Lambda'\right)\delta_{\Lambda \Lambda'-1}+\left(\kappa-\Lambda\right)\left(\kappa+\Lambda'\right)\delta_{\Lambda \Lambda'+1}+\left(\kappa+\Lambda\right)\left(\kappa+\Lambda'\right)\delta_{\Lambda 1-\Lambda'}\,.
\end{align}

\section{Comparison with short-distance regime matching for hybrid quarkonium}\label{shortcomp}

For quarkonium hybrids the matching in the short-distance regime, that is in the case $r\lesssim 1/\lQ$, of the static potential can be found in Refs.~\cite{Brambilla:1999xf,Berwein:2015vca,Brambilla:2019jfi}
\begin{align}
V_{1^+\Lambda}^{(0)}(\bm{r})=\Lambda_{1^+}+V^{(0)}_o(r)+b_{1^+\Lambda}r^2+\dots\label{a3eq2}
\end{align}
with $V^{(0)}_o(r)$ the perturbative octet potential. The constants $\Lambda_{1^+}$ and $b_{1^+\Lambda}$ are nonperturbative, local, gluon correlators. For example, the leading order of the static potential is given by the so-called gluelump mass, which takes the following form
\begin{align}
\Lambda_{1^+}&=\lim_{t\to\infty}\frac{i}{t}\ln
\,\langle 0|\bm{B}^{ia}(t/2,\,\bm{R}) \phi^{ab}(t/2,-t/2,\,\bm{R})\bm{B}^{ib}(-t/2,\,\bm{R})|0\rangle\,,
\end{align}
with $\phi^{ab}(t,t',\,\bm{R})$ the adjoint static Wilson line defined by Eq.~\eqref{stwl} with the gluon field in the adjoint representation.

The short-distance regime matching of the heavy quark spin-dependent potentials can be found in Ref.~\cite{Brambilla:2019jfi}. To compare our results with those of Ref.~\cite{Brambilla:2019jfi} we take heavy-quark spin-dependent potentials of Eq.~\eqref{sdp32} for the case $\kappa^{p}=1^+$ and use the Cartesian basis representation of the spin-1 matrices, i.e $\left(\bm{S}_1^i\right)^{jk}=-i\epsilon_{ijk}$. We find that
\begin{align}
&\Psi^{\dagger}_{1^+}\sum_{\Lambda\Lambda'}{\cal P}^{\rm c.r.}_{1\Lambda}\left[V^{sa}_{1^+\Lambda\Lambda'}(r)\bm{S}_{QQ}\cdot\left({\cal P}^{\rm c.r.}_{10}\cdot\bm{S}^{\rm c.r.}_{1}\right)+V^{sb}_{1^+\Lambda\Lambda'}(r)\bm{S}_{QQ}\cdot\left({\cal P}^{\rm c.r.}_{11}\cdot\bm{S}^{\rm c.r.}_{1}\right)\right]{\cal P}^{\rm c.r.}_{1\Lambda'}\Psi_{1^+}\nonumber\\
&=\left(\Psi^{\dagger}_{1^+}\right)^i\left[V^{sa}_{1^+11}(r)\bm{S}^k_{QQ}(\bm{S}^k_{1})_{ij}+\left(V^{sb}_{1^+10}(r)-V^{sa}_{1^+11}(r)\right)\bm{S}^k_{QQ}\left(\hat{\bm{r}}^i\hat{\bm{r}}^l(\bm{S}^k_{1})_{lj}-\hat{\bm{r}}^l(\bm{S}^k_{1})_{li}\hat{\bm{r}}^j\right)\right]\left(\Psi_{1^+}\right)^j\,.\label{a3eq1}
\end{align}
Using Eq.~\eqref{a3eq1} we can find the correspondence of our results to those of Ref.~\cite{Brambilla:2019jfi}
\begin{align}
&V^{sa}_{1^+11}(r)=V^{np\,(0)}_{SK}+V^{np\,(1)}_{SK}r^2\dots\\
&V^{sb}_{1^+10}(r)=V^{np\,(0)}_{SK}+\left(V^{np\,(0)}_{SKb}+V^{np\,(1)}_{SK}\right)r^2\dots
\end{align}
with $V^{np\,(0)}_{SK}$, $V^{np\,(1)}_{SK}$ and $V^{np\,(0)}_{SKb}$ nonperturbative, local, gluon correlators. For example, the leading-order one takes the form
\begin{align}
V^{np(0)}_{SK}&=\frac{c_F}{2}\lim_{T\to\infty}\frac{ie^{i\Lambda_{1^+} t}}{t}
\int^{t/2}_{-t/2}dt'\, \epsilon^{ijk}h^{bcd}\,\langle 0|\bm{B}^{ia}(t/2,\,\bm{R}) \phi^{ab}(t/2,\,t',\,\bm{R})g \bm{B}^{jc}(t,\,\bm{R})\phi^{de}(t/2,t',\,\bm{R}) \bm{B}^{ke}(-t/2,\,\bm{R})|0\rangle\,.\label{matchsk}
\end{align}

From Eqs.~\eqref{a3eq2} and \eqref{a3eq1} we can see the general structure for the short-distance expansion of the potentials: a nonanalitic, perturbative, term appears if the potential can be generated by interactions between the heavy quarks without involvement of the LDF, the gluons in the case of quarkonium hybrids. The nonperturbative contributions appear as coefficients of a series in powers of $r$. These coefficients are given by nonperturbative, local, gluon correlators. Furthermore, in the short-distance limit, $r\to 0$, the symmetry group $D_{\infty h}$ is enlarged to $O(3)\times C$ and therefore the potentials at leading order in the short-distance expansion are independent on the representation of $D_{\infty h}$ and only depend on the quantum numbers $\kappa^p$ of the LDF operators. The next-to-leading corrections are generated by $r$-dependent interactions of the heavy quarks with the LDF; these break the degeneracy of the $D_{\infty h}$ representations.

We can compare Eq.~\eqref{matchsk} to Eqs.~\eqref{pm1} and \eqref{pm2}. In both cases there are matching $c_F$ coefficients. Using Eq.~\eqref{a3eq2} we can see that the factor $ie^{i\Lambda_{1^+} t}/t$ in Eq.~\eqref{matchsk} is the short-distance expansion of the factor in front of the integral of Eqs.~\eqref{pm1}. The same holds true for the factor in front of the integral of Eq.~\eqref{pm2}, if we take into account the short-distance degeneracy of the $D_{\infty h}$ representations. One can also see that the $\bm{B}$ operators in Eq.~\eqref{matchsk} match those inserted in the Wilson loop of Eqs.~\eqref{pm1} and \eqref{pm2}: $\bm{B}(t/2,\,\bm{R})$ and $\bm{B}(-t/2,\,\bm{R})$ correspond to the operator in Eq.~\eqref{Qhybrid} inserted in the temporal sides of the Wilson loop and $\bm{B}(t,\,\bm{R})$ correspond to the NRQCD operator inserted in the spatial side of the Wilson loop.

\bibliographystyle{apsrev4-2}
\bibliography{biblioheh}

\end{document}